\documentclass[conference]{IEEEtran}
\IEEEoverridecommandlockouts

\usepackage{cite}
\usepackage{amsmath,amssymb,amsfonts}
\usepackage{algorithmic}
\usepackage{graphicx}
\usepackage{textcomp}
\usepackage{xcolor}
\usepackage{braket}
\usepackage{url}
\usepackage{subcaption}
\usepackage{bm}
\usepackage{multirow}
\usepackage{makecell}
\usepackage{siunitx}

\def\BibTeX{{\rm B\kern-.05em{\sc i\kern-.025em b}\kern-.08em
    T\kern-.1667em\lower.7ex\hbox{E}\kern-.125emX}}

\begin{document}
\bstctlcite{IEEEexample:BSTcontrol}

\title{Buffer-Qubit-Based Quantum Circuit Allocation Method Against Crosstalk Attacks}

\author{
  \IEEEauthorblockN{
    \parbox{\linewidth}{\centering
      Haruki~Nakayama\textsuperscript{*},
      Satoyuki~Tsukano\textsuperscript{†},
      Satoshi~Kono\textsuperscript{*},
      Keita~Matsumoto\textsuperscript{*},
      Toshio~Mori\textsuperscript{†},
      Takumi~Oikawa\textsuperscript{*},
      and Ryo~Uchida\textsuperscript{*}
    }
  }
  \vspace{2.0ex}
  \IEEEauthorblockA{
    \textsuperscript{*}Software Engineering Dept. 1, Systems Engineering Consultants Co.,LTD., Tokyo, Japan \\
    \{nakayama.haruki, kono.satoshi, matsumoto.keita, oikawa.takumi, uchida\}@sec.co.jp
  }
  \IEEEauthorblockA{
    \textsuperscript{†}Center for Quantum Information and Quantum Biology, The University of Osaka, Osaka, Japan \\
    \{tsukano.satoyuki, t.mori\}.qiqb@osaka-u.ac.jp
  }
}

\maketitle

\begin{abstract}
Quantum Multi-Programming (QMP) enables the simultaneous execution of quantum circuits from multiple users on a single quantum chip in cloud quantum computing services.
However, QMP is vulnerable to crosstalk attacks, in which a malicious user deliberately induces crosstalk to disrupt other users' computations.
As a countermeasure, prior work has proposed the use of buffer qubits, which remain idle and are allocated between quantum circuits to suppress the effect of crosstalk.
Nevertheless, to the best of our knowledge, such countermeasures have not been publicly reported in production quantum cloud software.
In this paper, we extend the QMP of OQTOPUS, open-source software that is already in production use for operating real quantum computers, with a buffer-qubit-based circuit allocation method.
Experiments on a real device characterize the attack effect in terms of energy relaxation and dephasing, showing that observable degradation appears in dephasing characteristics of qubits adjacent to the attacker circuit.
Based on this result, we release the proposed method as part of OQTOPUS and confirm that the buffer qubits are correctly reserved around the allocated circuits.
\end{abstract}


\section{INTRODUCTION}
The advent of cloud-based quantum computing services \cite{ibm_quantum, amazon_braket} has enabled researchers and developers worldwide to access quantum computers and perform quantum computations.
As the user base grows, efficient utilization of the limited computational resources has become increasingly important.
To address this issue, Quantum Multi-Programming (QMP), which allows quantum circuits from multiple users to be executed simultaneously on a single quantum chip, has been proposed and investigated \cite{quantum_mp}.
In QMP, since multiple users' quantum circuits are executed simultaneously on the same chip, crosstalk induced by one user's circuit can degrade the computation of other users  \cite{quantum_mp}.

Crosstalk attacks have been proposed as a security threat specific to QMP that exploits this vulnerability \cite{single_sided_crosstalk_attack, double_sided_crosstalk_attack}.
In this attack, a malicious user deliberately executes quantum circuits designed to induce crosstalk, thereby disrupting other users' circuits co-allocated on the same chip.
This threat has motivated research on mitigation techniques for crosstalk attacks in QMP.
The idea of using buffer qubits has been introduced to suppress crosstalk effects by keeping certain qubits idle between quantum circuits \cite{single_sided_crosstalk_attack, crosstalk_attack_countermeasures}.
Despite these efforts, to the best of our knowledge, such countermeasures have not been publicly reported in production quantum cloud software.

In this paper, we extend the existing QMP method in OQTOPUS \cite{oq_thesis}, a full-stack Open-Source Software (OSS) for cloud quantum computing that provides QMP \cite{oq_mp_auto}.
OQTOPUS has already been deployed as a cloud service and is used in the operation of quantum computing systems at The University of Osaka.
However, its current QMP implementation does not address crosstalk attacks.
To tackle this limitation, we propose a circuit allocation method that mitigates crosstalk attacks by incorporating buffer-qubit constraints.
Furthermore, we implement the proposed method in OQTOPUS and make it publicly available.

The main contributions of this paper are as follows:
\begin{itemize}
\item We propose a circuit allocation method for QMP that incorporates buffer-qubit constraints into a subgraph isomorphism problem formulation.
\item We characterize crosstalk attacks on a real superconducting quantum computer by measuring $T_1$ with $\ket{1}$ and $T_2^*$ with $\ket{+}$ and $\ket{i}$, showing that the observable degradation was confined to $T_2^*$ of qubits adjacent to the malicious circuit.
\item We implement and release the proposed method as a part of OQTOPUS, and verify that buffer qubits are correctly reserved in practice.
\end{itemize}

\section{THREAT MODEL}
\begin{figure}[]
  \centering
  \includegraphics[width=\columnwidth]{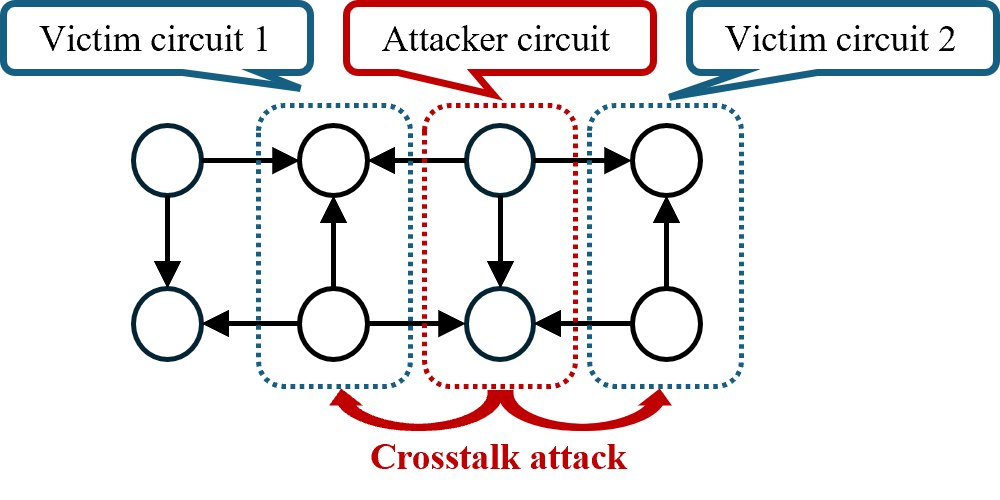}
  \caption{A schematic illustration of a crosstalk attack in QMP, depicting how an attacker circuit can affect victim circuits via crosstalk on an $8$-qubit device topology.}
  \label{fig:threat_model}
\end{figure}

Figure~\ref{fig:threat_model} illustrates an example scenario.
We assume a QMP environment in which the malicious user's quantum circuit and another user's quantum circuit can be executed simultaneously on the same chip.
In this setting, the scheduler, transpiler, and allocator are managed by the cloud service and are assumed to be trusted.
The malicious user is able to submit a quantum circuit consisting of CNOT gates, which are known to be a representative sources of crosstalk \cite{towards_antivirus_for_quantum_computers}.
This user's goal is to adversely affect the quantum computations of other users through the crosstalk.

\section{PROPOSED METHOD}
\begin{figure*}[]
  \centering
  \includegraphics[width=0.9\textwidth]{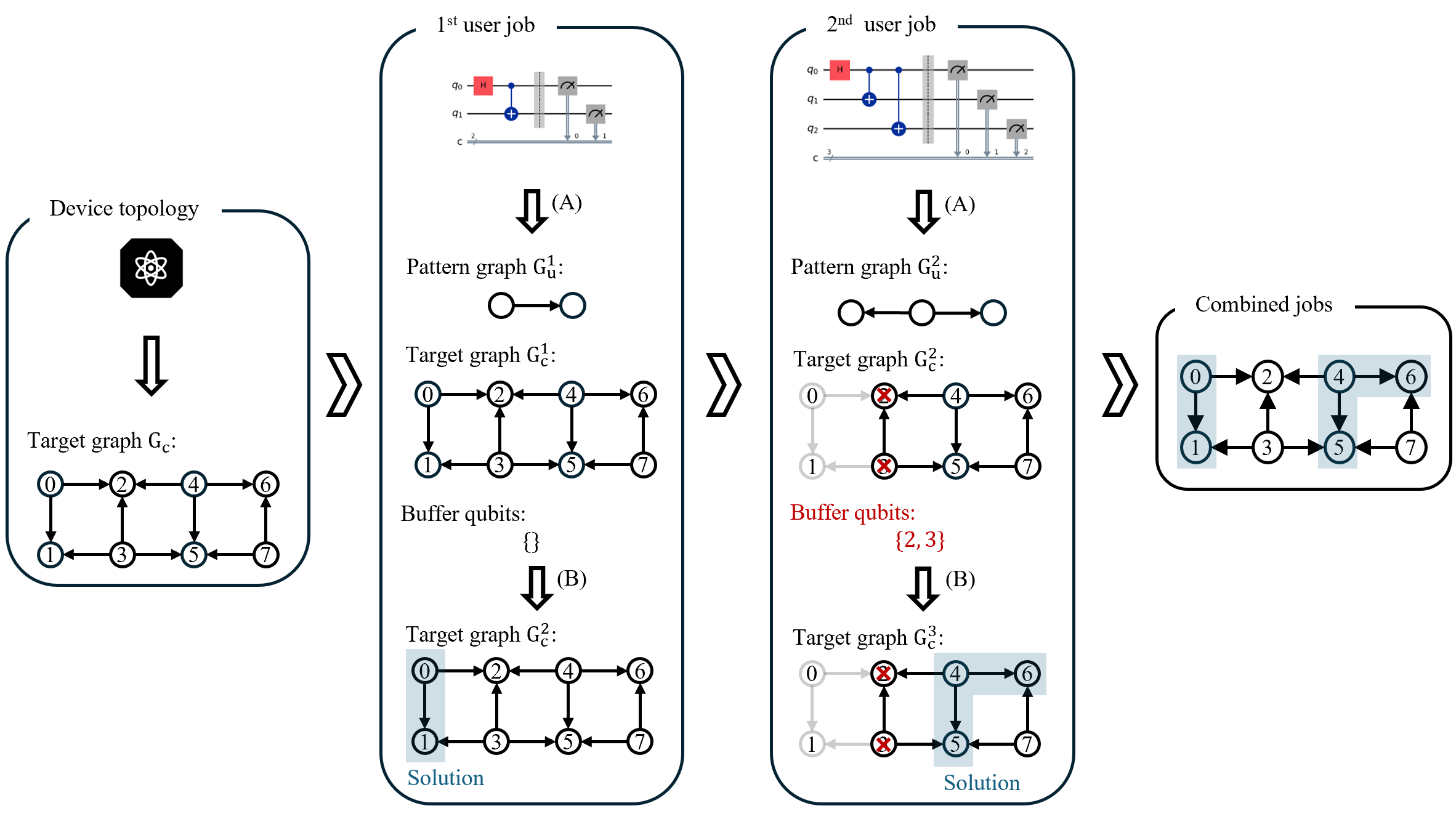}
  \caption{Illustration of the proposed circuit allocation procedure.
  For each user circuit, (A) a pattern graph $G_u^i$ is built from the transpiled circuit, and (B) $G_u^i$ is matched to a subgraph of the target graph $G_c^i$, subject to the constraint that vertices corresponding to buffer qubits in $V_b$ (marked with $\times$) are excluded from the mapping.
  After each allocation, unused qubits that fall within the crosstalk-affected range of the allocated circuit are added to $V_b$, and the procedure is repeated for the next circuit.
  The specific construction of $V_b$ depends on the range of crosstalk effects; in this example, qubits adjacent to the allocated ones are added to $V_b$ as one instance of such a construction.}
  \label{fig:proposed_method}
\end{figure*}

Figure~\ref{fig:proposed_method} shows an overview of the proposed method.
We extend the QMP method of OQTOPUS~\cite{oq_mp_auto}.
Specifically, we introduce a set $V_b$ of physical qubits that should be reserved as buffer qubits.
Multiple quantum circuits are then sequentially allocated on the same chip, subject to the constraints that physical qubits corresponding to vertices in $V_b$ cannot be used for allocation.
Since the construction of $V_b$ depends on the range of crosstalk attack effects, it is determined from experiments on the real quantum computer presented in Section~\ref{sec:crosstalk_attack_evaluation}.

The proposed method augments the Subgraph Isomorphism Problem (SIP) formulation used in the existing method with additional constraints, which prohibit the allocation of quantum circuits on physical qubits corresponding to buffer qubits.
By solving this constrained SIP sequentially, multiple quantum circuits are allocated onto the same chip.
SIP is the problem of deciding whether a pattern graph $G_1$ is isomorphic to a subgraph of a target graph $G_2$.

The notation used in the proposed method is summarized in Table~\ref{table:notation}.
A virtual qubit denotes an abstract qubit in a user's circuit before it is mapped to a physical qubit on the device.
The user's quantum circuit is assumed to be transpiled to match the device topology in advance, so $G_u^i$ is isomorphic to some subgraph of $G_c$.
However, $G_u^i$ is not necessarily isomorphic to a subgraph of $G_c^i$, because some physical qubits may already have been used or reserved as buffer qubits.
\begin{table}[b]
    \centering
    \caption{Notation used in the proposed method.}
    \label{table:notation}
    \begin{tabular}{ll}
        \hline
        Symbol & Description \\
        \hline
        $G_c$ & Graph $(V_c, E_c)$ representing the device topology. \\
        $V_c$ & Set of physical qubits on the device. \\
        $E_c$ & Set of directed edges representing pairs of\\
              & physical qubits on which CNOT gates can be executed. \\
        $G_c^i$ & Available target graph before allocating $G_u^i$ (with $G_c^1 = G_c$). \\
        \hline
        $G_u$ & Graph $(V_u, E_u)$ representing a user's quantum circuit. \\
        $V_u$ & Set of virtual qubits in the user's circuit. \\
        $E_u$ & Set of directed edges representing CNOT gates \\
              & executed on pairs of virtual qubits. \\
        $G_u^i$ & The $i$-th user circuit to be allocated. \\
        \hline
        $V_b$ & Set of physical qubits reserved as buffer qubits. ($V_b \subseteq V_c$)\\
        \hline
    \end{tabular}
\end{table}

Using the notation above, the detailed procedure of the proposed method is described as follows.
First, we solve the SIP with $G_u^i$ as the pattern graph and $G_c^i$ as the target graph, subject to the additional constraints that the image of the vertex mapping of $G_u^i$ does not include any vertex in $V_b$.
If $G_u^i$ is isomorphic to a subgraph of $G_c^i$, the virtual qubits corresponding to the vertices of $G_u^i$ are mapped to the physical qubits corresponding to the matched vertices in $G_c^i$.
Next, the vertices corresponding to the mapped physical qubits, together with their incident edges, are removed from $G_c^i$, and as a result, $G_c^{i+1}$ is obtained.
At the same time, unused physical qubits that fall within the crosstalk-affected range of the mapped circuit are added to $V_b$.
Then, the same procedure is repeated using the updated $G_c^{i+1}$, the updated $V_b$, and the next circuit $G_u^{i+1}$.

We formulate the above constrained SIP as a constraint programming problem and implement it in OQTOPUS~\cite{oq_implementation} using OR-Tools~\cite{cp-sat}.

\section{EVALUATION}
We conducted two experiments.
First, we evaluated the effects and spatial range of crosstalk attacks using a real superconducting quantum computer.
Second, on OQTOPUS with the proposed method implemented, we evaluated whether the qubits within the affected range observed in the previous experiment can be reserved as buffer qubits.

\subsection{Evaluation of Crosstalk Attack} \label{sec:crosstalk_attack_evaluation}
\begin{figure}[!h]
  \centering
  \includegraphics[width=\columnwidth]{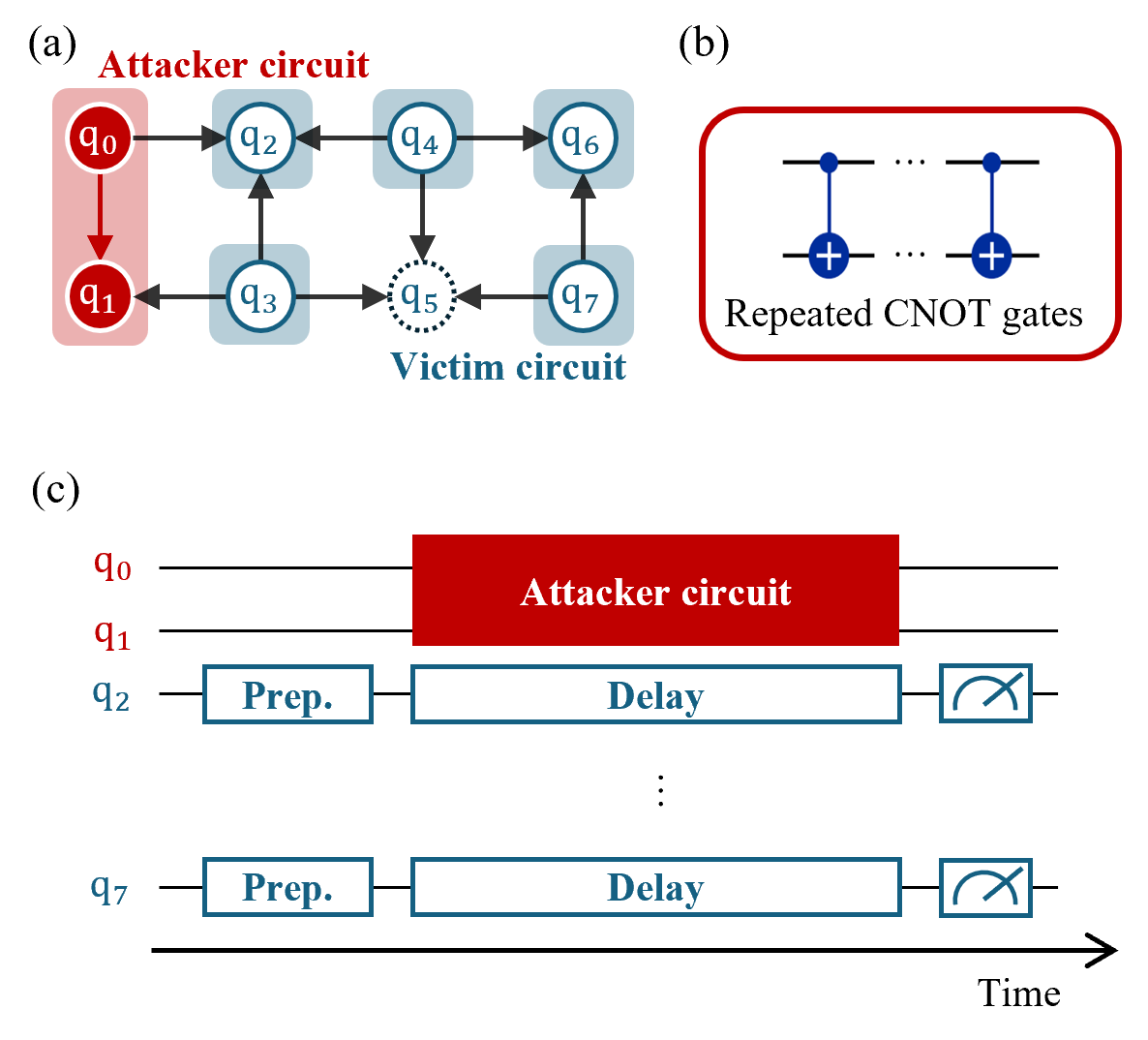}
  \caption{Overview of the experiment for evaluating the effect of crosstalk attacks.
    (a) Device topology and the allocation of each circuit.
    (b) Detailed view of the attacker circuit.
    (c) Illustration of the experimental procedure.
  }
  \label{fig:experiment_condition}
\end{figure}
An overview of the experiment is shown in Fig.~\ref{fig:experiment_condition}.
We evaluated the effects of crosstalk attacks on other quantum circuits executed simultaneously on the same chip, as well as the spatial range of these effects.
The experimental data were obtained using a quantum computer at The University of Osaka, whose qubit chip was provided by RIKEN.
In this experiment, we swept the number of repetitions of the CNOT gate in the attacker circuit and measured $T_1$ or $T_2^*$ of the qubits in victim circuits.
Since the elapsed time $t$ from state preparation to measurement in the victim circuit also increases as the number of CNOT gates increases, it is necessary to distinguish between the contribution from decoherence and that attributable to the crosstalk attack.
Therefore, as a reference experiment, we performed the same measurements under a condition in which the attacker circuit was not executed.
A delay was inserted so that the elapsed time matched that of the crosstalk attack condition.
The measurement conditions and the fitting functions are shown in Table~\ref{table:crosstalk_attack_conditions}.
For each data point, measurements with a shot count of $10{,}000$ were performed independently seven times.
The parameter values and their errors were then determined by fitting the obtained data.
For each qubit, we evaluated the difference between the parameter estimates under the attack and reference conditions using an approximate 95\% confidence interval, assuming independent fitting errors.
A difference was regarded as observable when the confidence interval excluded zero.

\begin{figure}[]
  \centering
    \includegraphics[width=0.95\linewidth]{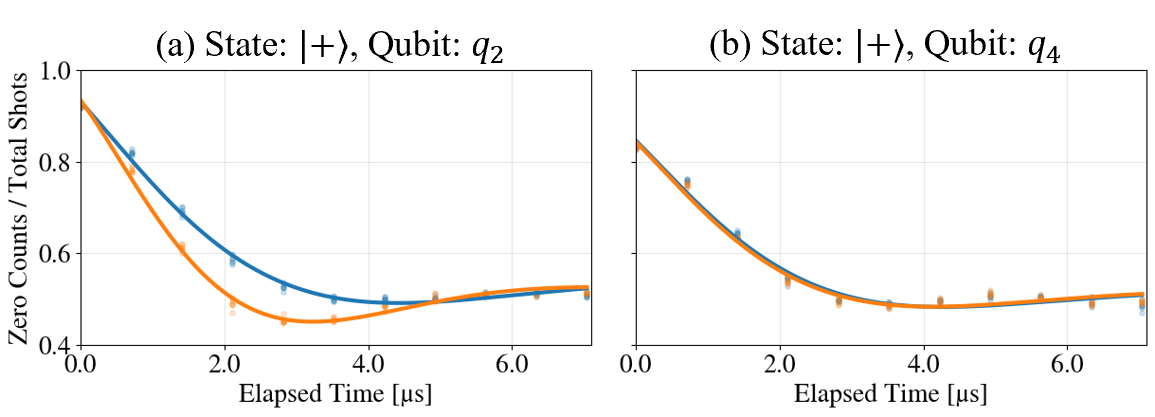}
    \caption{
      Zero-count fraction vs.\ elapsed time $t$ for the $\ket{+}$ state at (a) $q_2$ (distance $1$) and (b) $q_4$ (distance $2$).
      The blue and orange curves correspond to the reference and attack conditions, respectively.
      A separation appears only in (a).
    }
  \label{fig:fitting_results}
\end{figure}

Table~\ref{table:crosstalk_attack_results} shows that the approximate 95\% confidence intervals included zero for all $T_1$ measurements, whereas those for $T_2^\ast$ excluded zero only at qubits adjacent to the attacker circuit ($q_2$ for $\ket{+}$; $q_2$ and $q_3$ for $\ket{i}$).
This contrast is illustrated in Fig.~\ref{fig:fitting_results} for $\ket{+}$, with a separation at the adjacent qubit $q_2$ but not at the non-adjacent qubit $q_4$.
Accordingly, we add qubits adjacent to allocated circuits to $V_b$.

\begin{table}[]
    \centering
    \caption{Experimental conditions in the crosstalk attack evaluation.}
    \label{table:crosstalk_attack_conditions}
    \begin{tabular}{cccc}
        \hline
        \makecell{Prepared\\State} & \makecell{Measurement\\Axis} & \makecell{Fitting\\Function\textsuperscript{a}} & \makecell{\# of\\CNOT} \\
        \hline
        $\ket{1}$ & $Z$
        & $Ae^{-t/T_1}+B$
        & $0,5,\dots,50$ \\
        $\ket{+}$ & $X$
        & $Ae^{-t/T_2^*}\cos(\omega t)+B$
        & $0,1,\dots,10$ \\
        $\ket{i}$ & $Y$
        & $Ae^{-t/T_2^*}\cos(\omega t)+B$
        & $0,1,\dots,10$ \\
        \hline
    \end{tabular}

    \vspace{1mm}
    \footnotesize{\textsuperscript{a} $A$, $B$, $T_1$ (or $T_2^*$), and $\omega$ are the parameters. $t$ denotes the total elapsed time of the victim circuit.}
\end{table}
\begin{table}[]
    \centering
    \caption{Fitting results of $T_1$ and $T_2^*$. }
    \begin{tabular}{lllcc}
        \hline
        \makecell{Decoherence} & \makecell{State} & \makecell{Qubit \\ (Distance)} & Reference [\si{\micro\second}] & Attack [\si{\micro\second}] \\
        \hline
        $T_1$ & $\ket{1}$ & $q_2 (1)$ & $32.86\pm4.80$ & $25.41\pm2.84$ \\
        $T_1$ & $\ket{1}$ & $q_3 (1)$ & $15.76\pm2.32$ & $17.68\pm2.42$ \\
        $T_1$ & $\ket{1}$ & $q_4 (2)$ & $26.89\pm2.05$ & $26.16\pm1.84$ \\
        $T_1$ & $\ket{1}$ & $q_6 (3)$ & $32.59\pm3.84$ & $28.83\pm3.09$ \\
        $T_1$ & $\ket{1}$ & $q_7 (3)$ & $12.25\pm4.98$ & $9.93\pm3.10$ \\
        \hline
        \textbf{$T_2 ^ *$} & \textbf{$\ket{+}$} & \textbf{$q_2 (1)$} & $\bm{2.28\pm0.07}$ & $\bm{1.95\pm0.05}$ \\
        $T_2 ^ *$ & $\ket{+}$ & $q_3 (1)$ & $2.53\pm0.12$ & $2.44\pm0.09$ \\
        $T_2 ^ *$ & $\ket{+}$ & $q_4 (2)$ & $2.10\pm0.12$ & $2.03\pm0.10$ \\
        $T_2 ^ *$ & $\ket{+}$ & $q_6 (3)$ & $2.35\pm0.12$ & $2.33\pm0.11$ \\
        $T_2 ^ *$ & $\ket{+}$ & $q_7 (3)$ & $2.23\pm0.14$ & $2.09\pm0.15$ \\
        \hline
        \textbf{$T_2 ^ *$} & \textbf{$\ket{i}$} & \textbf{$q_2 (1)$} & $\bm{2.50\pm0.09}$ & $\bm{2.11\pm0.07}$ \\
        \textbf{$T_2 ^ *$} & \textbf{$\ket{i}$} & \textbf{$q_3 (1)$} & $\bm{3.00\pm0.16}$ & $\bm{2.61\pm0.10}$ \\
        $T_2 ^ *$ & $\ket{i}$ & $q_4 (2)$ & $2.09\pm0.11$ & $1.98\pm0.11$ \\
        $T_2 ^ *$ & $\ket{i}$ & $q_6 (3)$ & $2.49\pm0.12$ & $2.29\pm0.12$ \\
        $T_2 ^ *$ & $\ket{i}$ & $q_7 (3)$ & $2.20\pm0.22$ & $1.83\pm0.18$ \\
        \hline
    \end{tabular}

    \vspace{1mm}
    \footnotesize{Boldface entries indicate qubits for which the approximate 95\% confidence interval for the difference between the reference and attack estimates excludes zero.These values correspond to the fitted parameter estimates, with uncertainties representing the fitting errors.}
    \label{table:crosstalk_attack_results}
\end{table}

\subsection{Evaluation of Proposed Method}
Using the construction of $V_b$ determined in Section~\ref{sec:crosstalk_attack_evaluation}, we confirmed that the proposed method correctly reserves buffer qubits between circuits in OQTOPUS.

\begin{figure}
  \centering
  \includegraphics[width=0.71041\columnwidth]{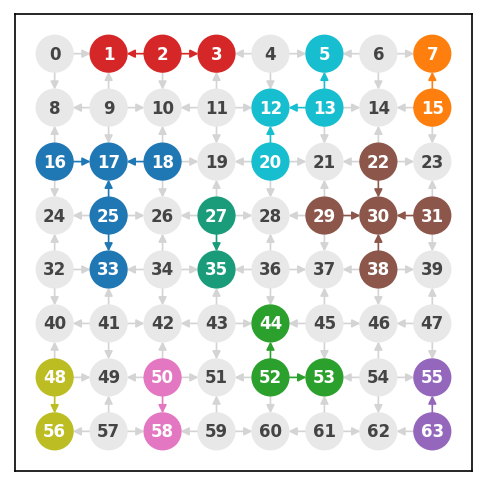}
  \caption{
    Quantum circuit allocation for a single batch, in which $30$ colored qubits were allocated to user jobs and $34$ grayed qubits were reserved as buffer qubits.
  }
  \label{fig:allocation_result}
\end{figure}

Randomly generated $100$ quantum circuits consisting of $2$ to $5$ qubits were processed using QMP with the proposed method on a device topology in which $64$ qubits were arranged on a square lattice.
By partitioning these quantum circuits into $12$ execution batches, the proposed method successfully allocated all circuits without violating the buffer-qubit constraints (Fig.~\ref{fig:allocation_result}).
Although buffer qubits accounted for 53.1\% of the chip in this case, this fraction may decrease as the number of qubits used by each user circuit increases.
Quantitatively evaluating the resulting throughput overhead under different circuit-size distributions remains an important direction for future work.

\section{CONCLUSION}
In this paper, we proposed a circuit allocation method for mitigating crosstalk attacks in QMP. The method extends the existing QMP in OQTOPUS by introducing buffer-qubit constraints that are dynamically updated during qubit allocation.
Our experiments on a real superconducting quantum computer showed that, under our experimental conditions, crosstalk effects were limited to qubits adjacent to the attacker circuits.
Based on this observation, we implemented the proposed method in OQTOPUS and released it as OSS.
We demonstrated that circuit allocation avoids qubits within the affected range and that buffer qubits are correctly inserted in practice.
Future work includes the development of a quantitative method for evaluating throughput, end-to-end evaluation of the mitigation effect on victim-circuit outputs, and evaluations against crosstalk attacks with longer-range effects~\cite{qubithammerremotely}.

\section*{Acknowledgment}
This work was supported by JST COI-NEXT Program, Grant No. JPMJPF2014.

\addcontentsline{toc}{section}{Acknowledgment}

\bibliographystyle{IEEEtran}
\bibliography{ref}

\end{document}